\documentclass[pdflatex,sn-mathphys-num]{sn-jnl}

\usepackage{graphicx}%
\usepackage{multirow}%
\usepackage{amsmath,amssymb,amsfonts}%
\usepackage{amsthm}%
\usepackage{mathrsfs}%
\usepackage[title]{appendix}%
\usepackage{xcolor}%
\usepackage{textcomp}%
\usepackage{manyfoot}%
\usepackage{booktabs}%
\usepackage{algorithm}%
\usepackage{algorithmicx}%
\usepackage{algpseudocode}%
\usepackage{listings}%
\usepackage{natbib}%
\usepackage{blindtext}
\usepackage{enumitem}
\usepackage{makeidx}
\usepackage{mathdots}
\usepackage{graphicx}
\usepackage{hyperref}
\usepackage{multirow}
\usepackage{tabularx}
\usepackage{nicematrix}
\usepackage{booktabs}
\usepackage{changepage}
\usepackage{makecell}
\usepackage{array}

\theoremstyle{thmstyleone}%
\theoremstyle{thmstyletwo}%

\theoremstyle{thmstylethree}%

\begin{document}

\title[Article Title]{Studying the Behavior of Radial Free Geodesics in $\Lambda$CDM Model}


\author*[1]{\fnm{Omar} \sur{Nemoul}}\email{omar.nemoul@umc.edu.dz}

\author[1]{\fnm{Hichem} \sur{Guergouri}}\email{hichem.guergouri@univ-bejaia.dz}

\author[1,2]{\fnm{Jamal} \sur{Mimouni}}\email{jamalmimouni@umc.edu.dz}

\affil*[1]{\orgdiv{Research Unit in Scientific Mediation}, \orgname{CERIST}, \orgaddress{\city{Constantine}, \postcode{25016}, \country{Algeria}}}

\affil[2]{\orgdiv{Laboratoire de Physique Mathematique et Subatomique}, \orgname{University of Constantine 1}, \orgaddress{\city{Constantine}, \postcode{25017}, \country{Algeria}}}


\abstract{This paper presents an analytical study of the behavior of radial free-geodesics in the Friedmann-Lemaître-Robertson-Walker (FLRW) spacetime within the Lambda Cold Dark Matter ($\Lambda$CDM) model.  Using the radial free motion solutions, we provide two methods for characterizing the geodesics and defines a general formula that encapsulates all possible solutions, determined by two initial conditions. We show that the past light cone, event horizon, and particle horizon, can be considered as special cases of this overarching formula. Furthermore, the paper explores the free geodesics within the currently accepted cosmological model based on the recent Planck results, thoroughly examining the various possible geodesic scenarios.}




\maketitle

\section{Introduction}\label{sec1}
Experiments exploring the Cosmic Microwave Background (CMB), such as COBE \cite{bib1}, WMAP \cite{bib2}, and Planck \cite{bib3}, have provided substantial evidence supporting the $\Lambda$CDM model \cite{bib4}. These comprehensive studies have consistently affirmed that, at large scales, the observable universe exhibits spatial homogeneity, isotropy, and flatness. These characteristics crucially validate the FLRW metric \cite{bib5,bib6,bib7,bib8} as the framework for describing our spacetime \cite{bib9}. The concept of geodesics holds significant relevance not only in cosmology but also in astrophysics and quantum gravity. The study of geodesics, with its extensive applications, has seen significant analytical efforts aimed at solving their equations of motion. Research on geodesic motions in FLRW spacetimes, particularly for radially freely-falling test particles in the absence of non-gravitational forces, has been extensively explored in Refs. \cite{bib10, bib11, bib12, bib13, bib14, bib15, bib16}. Whiting's work \cite{bib10} signaled a significant early contribution to this field. He derived the equations of motion for a free particle within a Newtonian framework and then extended these concepts to a relativistic context. Subsequently, Ref. \cite{bib11} investigated geodesic motions in the low peculiar velocity regime, asking the critical question: does a test galaxy approach, recede, or stay at the same distance relative to a given reference frame? This inquiry was specifically discussed within the context of an accelerating universe characterized by $(\Omega_m,\Omega_\Lambda )=(0.3,0.7)$. Our analysis in this paper will address this question for all peculiar velocity regimes and the parameters of the recent Planck results \cite{bib3} $(\Omega_m,\Omega_\Lambda )=(0.315,0.685)$. It is worth noting that the approaches in Refs. \cite{bib10, bib11} exhibited several limitations in calculation and interpretation for solving geodesic motion. Addressing these limitations, Ref. \cite{bib12} presented a general solution for geodesics within the framework of general relativity, including an examination of a universe with a single dominant component. Note that, Ref. \cite{bib10} claimed that a freely-falling particle moving uniformly in an expanding universe would eventually converge with the Hubble flow. However, this assertion was challenged and refuted in Ref. \cite{bib13}, where the authors elaborated on seven specific conditions necessary for a particle to asymptotically merge the Hubble flow. They formally demonstrated that particles following general geodesic paths do not always asymptotically rejoin the Hubble flow in all eternally expanding universe. A necessary condition for a particle to asymptotically reach the Hubble flow, as outlined in their rigorous analysis, is contingent upon the satisfaction of all seven conditions, particularly emphasizing the crucial criterion $w_d < -\frac{1}{3}$, where $w_d$ represents the equation of state parameter for the dominant cosmic component as time approaches infinity. For recent research, Ref. \cite{bib14} used conformal time transformations to derive a general analytical formulation. This study provided insights into various cosmological models, including the Flat FLRW spacetime without a cosmological constant, the Milne universe, the de-Sitter Universe, and the Anti-de Sitter universe, and it also explored scenarios involving return journeys. In another notable work \cite{bib15}, the Pseudo-Painlevé-Gullstrand coordinates were used to solve the geodesic equations, investigating cases of a single dominant component in flat FLRW, radiation-dominant universe, free scalar field-dominant universe, and the de Sitter universe. All these investigations \cite{bib10, bib12, bib13, bib14, bib15} primarily focused on solving the geodesic equations. A particularly elegant approach was presented in Ref. \cite{bib16} by Cotăescu, who introduced a method to determine the dynamics of a freely moving test particle without directly dealing with the geodesic equations. Using the Killing vectors associated with the $E(3)$ isometry in flat FLRW spacetime, Cotăescu successfully extracted a conserved quantity which can be used to determine the geodesics.
\\

Here in this work, we shall use the geodesic solutions derived in \cite{bib12, bib13, bib14, bib15, bib16} for a radially freely-falling particle to study the behavior of geodesic free motion within the framework of the currently accepted cosmological model \cite{bib3}. In the beginning, our next two sections offer an in-depth overview of free motion in FLRW spacetime, we will determine the timelike geodesics for radial motion of a test particle in FLRW spacetime by using directly the stationary-action principle \cite{bib17}\footnote{This approach is commonly presented in nearly all introductory texts on general relativity (e.g., Hartle, Gravity: An introduction to Einstein's General Relativity, 1st edition, Chapter 8).}, without solving the geodesic equations—a task that typically requires the explicit computation of Christoffel symbols and the resolution of two independent differential equations. Subsequently, we introduce two distinct yet equivalent methods for characterizing the physical solutions. The first parametrization, designated as $(\chi_i, v_i)$, involves the initial comoving radial distance and the initial peculiar velocity at the initial time $t_i$. The second parametrization, represented as $(\chi_i, v_0)$, considers the initial comoving radial distance at initial time $t_i$ and the peculiar velocity at the present time $t_0$. These parametrizations facilitate a comprehensive analysis of the motion under consideration. For reasons we will delve into later, we choose the second parametrization $(\chi_i, v_0)$ for a detailed study of a free-falling particle, starting from the Big Bang singularity $(i=\mathrm{BB})$ at initial time $t_i=0$. This approach will facilitate in-depth exploration into the dynamics of free geodesics. Using these specified initial conditions, we shall formulate a general expression for the physical (proper) radial distance. We will illustrate how this formula can be applied to analyze specific cases such as the past light cone, event horizon, and particle horizon. Furthermore, our results will be applied within the currently accepted cosmological model ($\Lambda$CDM), using the recent Planck results \cite{bib3}. The outcomes will be visualized through a series of graphs in both comoving and physical frames. By using the physical velocity and acceleration, our analysis will further explore various scenarios of a free-falling traveler relative to a comoving reference frame. This will involve determining the initial conditions $(\chi_i, v_0)$ that correspond to both a one-way journey and a return journey with its three possibilities, each of which will be detailed in the subsequent sections of our paper. Additionally, within this framework, we introduce and examine the concept of "access conditions" for showing that no free traveler can remain at a constant physical distance relative to a comoving observer. Finally, we will finish the discussion by drawing our conclusions. For the remainder of this article, we will use the Greek letters $\mu,\nu,...$ assigned values 0,1,2, and 3 to signify spacetime indices. Spatial indices will be denoted using the Latin letters $i,j,...$ with values 1,2, and 3. Our analysis will use a metric of a signature $(+,-,-,-)$ in a spacetime coordinate system defined by the cosmic time $t$ and comoving spatial coordinates $x^i$. The notation of a dot placed above a variable, as in $\dot{x}^i$ denotes its time derivative. We define $t_0$ as the present time, and the cosmological scale factor now will be normalized to one $a(t_0 )=1$. Additionally, we will use a system of units in which the speed of light is $c=1$.
\section{Kinematic Concepts}\label{sec2}
Before turning to address geodesics for free motion, it is essential to discuss various kinematic concepts fundamental to our study. The FLRW spacetime serves a crucial mathematical framework for describing the large-scale structure and evolution dynamics of the universe. This model assumes that, on average, the universe is both homogeneous (similar at all points) and isotropic (looks the same in all directions) on large cosmic scales. It provides a powerful tool for understanding the expansion of the universe and its fundamental characteristics. Its spacetime interval can be straightforwardly written as follows
\begin{equation}
ds^2 = g_{\mu\nu}dx^\mu dx^\nu = dt^2 - a^2(t)\gamma_{ij}(\vec{x})dx^i dx^j
\label{eq1},
\end{equation}
where $a(t)$ is the scale factor that represents the expansion of the 3d-space, $\gamma_{ij} (\vec{x})$ is the spatial homogeneous and isotropic "comoving" 3d- metric in the comoving coordinates system $(x^i)$, while the physical 3d- metric is $a^2(t)\gamma_{ij} (\vec{x})$. Consider a test particle that follows a specific timelike path in the FLRW spacetime. Its equation of motion in the comoving coordinate $(x^i)$
\begin{subequations}
\begin{align}
\text{comoving position:} \ \ \quad t \mapsto x^{i}(t)
\label{eq2a}\\
\text{comoving velocity:} \ \  \quad t \mapsto \dot{x}^{i}(t)
\label{eq2b}
\end{align}
\end{subequations}
The equation of motion in the physical coordinate $(x_{\mathrm{phy}}^i=a(t) x^i)$:
\begin{subequations}
\begin{align}
\text{physical position:} \ &t \mapsto x_{\mathrm{phy}}^i(t) = a(t)x^i(t) \label{eq3a}\\
\text{physical velocity:} \ &t \mapsto \dot{x}_{\mathrm{phy}}^i(t) = a(t)\dot{x}^i(t) + H(t)x_{\mathrm{phy}}^i(t) \label{eq3b}
\end{align}
\end{subequations}
We can decompose the physical velocity into a peculiar (proper velocity) $v_{\text{pec}}^i = a(t) \dot{x}^i(t)$ and a recessional (Hubble flow) $v_{\text{rec}}^i = H(t) x_{\text{phy}}^i(t)$. Now the comoving covariant 4-velocity $V^\mu$ is defined to be the tangent vectors along the path with respect to the proper time $\tau$. It is given by
\begin{equation}
V^\mu = \frac{dx^\mu}{d\tau}
\label{eq4}.
\end{equation}
Proper time is the measured time in the proper reference frame of the test particle, which is the frame $(t_\mathrm{p} \equiv \tau, x_\mathrm{p}^i)$ in which the test particle is “always” at rest, that is
\begin{equation}
dx_\mathrm{p}^i = 0
\label{eq5}.
\end{equation}
Here the spacetime interval made by the motion of the test particle as measured in its own proper reference frame is given by
\begin{equation}
ds^2 = d\tau^2
\label{eq6},
\end{equation}
and since $ds^2$ is an invariant quantity (does not depend on the choice of the reference frame). Therefore, by using Eqs. \eqref{eq1} and \eqref{eq6}, one can express the proper time in terms of the cosmic time for a timelike traveling test particle as follows
\begin{equation}
 d\tau = dt \sqrt{1 - a^2(t) \gamma_{ij}(\vec{x}) \dot{x}^i \dot{x}^j} = dt \sqrt{1 - |\vec{v}_{\text{pec}}|^2}
\label{eq7}
\end{equation}
$|\vec{v}_{\text{pec}}(t)|$ is the magnitude square of the peculiar velocity and it is written by
\begin{equation}
|\vec{v}_{\text{pec}}(t)|^2 = a^2(t)\gamma_{ij}(\vec{x}) \dot{x}^i(t) \dot{x}^j(t)
\label{eq8}.
\end{equation}
Consequently, the comoving 4- velocity can be expressed through the time derivative as follows
\begin{equation}
V^\mu = \frac{dt}{d\tau} \frac{dx^\mu}{dt} = (\gamma(t,\vec{x}), \gamma(t,\vec{x}) \dot{x}^i(t))
\label{eq9},
\end{equation}
where the Lorentz factor $\gamma$ can be determined from Eq. \eqref{eq7} and it depends on the scale factor as
\begin{equation}
\gamma(t) = \frac{dt}{d\tau}= \frac{1}{\sqrt{1 - a(t)^2 \gamma_{ij}(\vec{x}) \dot{x}^i(t) \dot{x}^j(t)}}
 = \frac{1}{\sqrt{1 - |\vec{v}_{\text{pec}}|^2}}
\label{eq10}.
\end{equation}
We must not confuse between the Lorentz factor $\gamma$ and the spatial comoving metric $\gamma_{ij}(\vec{x})$. It is trivial to show that from Eqs. \eqref{eq1}, \eqref{eq4} and \eqref{eq6}, the comoving 4- velocity vector is always normalized to one as
\begin{equation}
g_{\mu\nu} V^\mu V^\nu = \frac{ds^2}{d\tau^2} = 1
\label{eq11}
\end{equation}
The comoving covariant 4- momentum $P^\mu$ of a mass $m$ moving along a path with a comoving 4-velocity vector $V^\mu$ is defined as follows
\begin{equation}
P^\mu = mV^\mu = (m\gamma, m\gamma \dot{x}^i)
\label{eq12}.
\end{equation}
From Eqs. \eqref{eq11} and \eqref{eq12}, we can see
\begin{equation}
g_{\mu\nu} P^\mu P^\nu = m^2
\label{eq13},
\end{equation}
which can be explicitly written as,
\begin{equation}
(P^0)^2 - a^2(t) \gamma_{ij}(\vec{x}) P^i P^j = m^2
\label{eq14}.
\end{equation}
The 0$^\text{th}$ component (time component) of this 4- momentum vector represents the physical energy $P^t = E = m\gamma$ of the test particle. To obtain the energy-momentum relation $E^2 = p^2 + m^2$ (mass-shell condition), the physical 3- momentum $p^i$ should be expressed in terms of the comoving 3- momentum $P^i$ as follows
\begin{equation}
p^i =  a(t) P^i 
\label{eq15}.
\end{equation}
Now, let's verify that the expression \eqref{eq12} indeed corresponds to the comoving 4-momentum of a test particle in FLRW spacetime. This verification is based on the principle that physical motion follows the shortest path in spacetime. This is determined by minimizing the spacetime interval $\Delta s$ (the proper time interval) and therefore, we define an action principle as
\begin{equation}
S[x^i(t)] = -m \int ds = -m \int dt \sqrt{1 - a^2(t) \gamma_{ij}(\vec{x}) \dot{x}^i \dot{x}^j}
\label{eq16},
\end{equation}
where the Lagrangian is given by
\begin{equation}
L(x^i(t), \dot{x}^i(t), t) = -m\sqrt{1 - a^2(t) \gamma_{ij}(\vec{x}) \dot{x}^i \dot{x}^j}
\label{eq17}.
\end{equation}
To ensure that the Lagrangian is expressed in energy units and to obtain the correct nonrelativistic Newton's law, we have multiplied by the factor “$-m$”. The comoving canonical 3-momentum $P_i$ corresponds to $x^i$, given by
\begin{equation}
P_i = \frac{\partial L}{\partial \dot{x}^i} = \frac{ma^2(t) \gamma_{ij}(\vec{x}) \dot{x}^j}{\sqrt{1 - a^2(t) \gamma_{ij}(\vec{x}) \dot{x}^i \dot{x}^j}} = ma^2(t) \gamma \dot{x}_i
\label{eq18},
\end{equation}
with $P^i=\frac{\gamma^{ij}}{a^2(t)}P_j$ as proved earlier in Eq. \eqref{eq12}. For the $0^{\mathrm{th}}$ component $P^t = P_t$, which represents the energy of the test particle. This can be checked by using the Legendre transform of the Lagrangian $L$ to the energy $E$ as follows
\begin{equation}
E = P_i \dot{x}^i - L = \frac{m}{\sqrt{1 - a^2(t) \gamma_{ij}(\vec{x}) \dot{x}^i \dot{x}^j}} = m\gamma
\label{eq19},
\end{equation}
which is the same formula in Eq. \eqref{eq12}. Thus, the comoving covariant 4- momentum, as detailed earlier, is thoroughly checked. Now, to determine the geodesics, we apply the Euler-Lagrange equation to extremize the action \eqref{eq16}
\begin{equation}
\frac{d}{dt} \left( \frac{\partial L}{\partial \dot{x}^i} \right) = \frac{\partial L}{\partial x^i}
\label{eq:20}.
\end{equation}
Applying these equations yields three second order differential equations for the three spatial comoving coordinates $x^i (t)$
\begin{equation}
\frac{d}{dt} \left( \frac{a^2(t) \gamma_{ij}(\vec{x}) \dot{x}^j}{\sqrt{1 - a^2(t) \gamma_{i'j'}(\vec{x}) \dot{x}^{i'} \dot{x}^{j'}}} \right) = \frac{a^2(t) \dot{x}^k \dot{x}^l\left( \frac{\partial \gamma_{kl}(\vec{x})}{\partial x^i} \right)}{2\sqrt{1 - a^2(t) \gamma_{i'j'}(\vec{x}) \dot{x}^{i'} \dot{x}^{j'}}}
\label{eq21}
\end{equation}
It can be proved that the set of these three differential equations \eqref{eq21}, when combined with the expression for $\frac{dt}{d\tau}$ as indicated in Eq. \eqref{eq10}, is equivalent to the system of four geodesic equations
\begin{equation}
\frac{d^2 x^\mu}{ds^2} + \Gamma_{\alpha\beta}^\mu \frac{dx^\alpha}{ds} \frac{dx^\beta}{ds} = 0
\label{eq22},
\end{equation}
for the FLRW metric \eqref{eq1}. Solving these equations \eqref{eq21} to find general geodesics is a complex task. Therefore, for simplicity and without losing generality, we will focus on radial geodesics in the following section.
\section{Radial Geodesic Motion in FLRW}\label{sec3}
Now, we will determine the timelike geodesics of a freely-falling test particle within the FLRW spacetime, using the comoving coordinates. Assuming that our universe is characterized by a homogeneous and isotropic space where no direction or location are preferred, any general geodesic curve can be transformed to have a purely radial spatial part through an appropriate choice of local coordinates. Hence, without loss of generality, radial geodesics will be considered in what follows. In the comoving spherical coordinates system $x^i = (\chi, \theta, \phi)$, we consider $\theta$ and $\phi$ as constants, while the comoving radial coordinate $\chi = \chi(t)$ varies with time. Consequently, this leads to the induced spacetime interval \eqref{eq1} for this radial motion in the following form
\begin{equation}
ds^2 = dt^2 - a^2(t) d\chi^2
\label{eq23}.
\end{equation}
The angular component of the comoving 4-momentum vanishes, i.e., $P^\theta = P^\phi = 0$, the radial peculiar velocity will have the form $v_{\text{pec}}^i(t) = (a(t) \dot{\chi}(t), 0, 0)$ with a magnitude $v_{\text{pec}} \equiv |\vec{v}_{\text{pec}}| = a(t) \dot{\chi}(t)$ and the comoving 4- momentum is now time dependent only and can be written as
\begin{equation}
P^\mu(t) = (m\gamma(t), m\gamma(t) \dot{\chi}(t), 0, 0)
\label{eq24},
\end{equation}
with the Lorentz factor
\begin{equation}
\gamma(t) = \frac{dt}{d\tau} = \frac{1}{\sqrt{1 - a^2(t) \dot{\chi}^2(t)}} = \frac{1}{\sqrt{1 - v_{\text{pec}}^2(t)}}
\label{eq25}.
\end{equation}
The physical energy and 3- momentum are given by
\begin{subequations}
\begin{align}
E(t) = P^t = \frac{m}{\sqrt{1 - v_{\text{pec}}^2(t)}}
\label{eq26b}\\
p(t) =a(t) P^\chi = \frac{m v_{\text{pec}}}{\sqrt{1 - v_{\text{pec}}^2(t)}}
\label{eq26a}.
\end{align}
\end{subequations}
\subsection{From Stationary-Action Principle}\label{subsec.3.1}
A geodesic is defined as a curve whose tangent vectors remain parallel to themselves when transported along the curve. Alternatively, it is a curve that minimizes the proper time $\Delta \tau$ (the spacetime interval $\Delta s$) between two spacetime points. Typically, geodesics are determined by solving the geodesic equations \eqref{eq22}. This process, in turn, can be quite complex, stemming from the need to explicitly calculate the Christoffel symbols $\Gamma^\mu_{\alpha\beta}$ and subsequently solve the independent differential equations. The primary challenge in handling the geodesic differential equation arises from the assumption that the general Lagrangian depends on the coordinates and their time derivatives. However, in the following section, we will bypass this intricate approach by directly applying Euler-Lagrange equation for the relevant variable $\chi(t)$. For our purposes, since we are focused on radial motion, the spacetime interval element in Eq. \eqref{eq23} is only depending on the time derivative of the radial coordinate, as follows
\begin{equation}
ds = dt \sqrt{1 - a^2(t) \dot{\chi}^2(t)}
\label{eq27}.
\end{equation} 
Consequently, although the geodesic equations are dervied from the least action principle, but addressing the problem directly by minimizing the spacetime interval is much simpler and more straightforward compared to engaging with the geodesic equations (for more details see \cite{bib17}). Therefore, we define the action $S[\chi(t)]$ for the radially freely-falling test particle in the FLRW spacetime as follows
\begin{equation}
S[\chi(t)] = -m \int ds = -m \int dt \sqrt{1 - a^2(t) \dot{\chi}^2(t)}
\label{eq28}.
\end{equation}
The Lagrangian for this action is given by
\begin{equation}
L(\dot{\chi},t) = -m\sqrt{1 - a^2(t) \dot{\chi}^2(t)}
\label{eq29}.
\end{equation}
It is evident that this Lagrangian is symmetric under any global (time-independent) translation $\epsilon$ of the comoving radial variable $\chi$, as the Lagrangian is independent of $\chi$
\begin{equation}
\chi \rightarrow \chi' = \chi + \epsilon
\label{eq30}.
\end{equation}
Consequently, according to Emmy Noether's theorem, this dynamical system obeys a conservation law, represented by a conserved quantity (constant of motion)\footnote{This conserved quantity is represented in numerous papers and textbooks by the Killing vector associated with the translational symmetry of the radial distance, see for instance \cite{bib16} and \cite{bib17}.}, which remains invariant over time. The constant of motion can be directly obtained from the Euler-Lagrange equation \eqref{eq:20}; it yields
\begin{equation}
\frac{d}{dt} \frac{\partial L}{\partial \dot{\chi}} = 0
\label{eq31}.
\end{equation}
From this equation, we can deduce that the conjugate momentum of $\chi$, denoted as $P_\chi$, is the corresponding conserved quantity associated with the global translation symmetry, as indicated by
\begin{align}
P_\chi&=\frac{\partial L}{\partial \dot{\chi}}=\frac{ma^2(t)\dot{\chi}(t)}{\sqrt{1-a^2(t)\dot{\chi}^2(t)}}=\frac{ma(t)v_{\text{pec}}(t)}{\sqrt{1-v_{\text{pec}}^2(t)}}\nonumber\\
&=a(t)p(t)= a^2(t)m \frac{dx^\chi}{d\tau} = a^2(t)P^\chi
\label{eq32},
\end{align}
with
\begin{equation}
\frac{dP_\chi}{dt} = 0
\label{eq33}.
\end{equation}
Eq. \eqref{eq32} is consistent with the result obtained in Ref. \cite{bib16}. Considering the constant of motion $P_\chi = mA$, where $A$ is an arbitrary real number. Inverting Eq. \eqref{eq32} straightforwardly allows us to derive the comoving radial velocity $\dot{\chi}$, expressed as
\begin{equation}
\dot{\chi}(t) = \frac{1}{a(t)}\frac{A}{\sqrt{a^2(t) + A^2}}
\label{eq34}.
\end{equation}
The real constant $A$ is related to the initial velocity of the particle. This expression in Eq. \eqref{eq34} is consistent with the results derived in Refs. \cite{bib11, bib12, bib13, bib14, bib15, bib16}. Indeed, the authors in Ref. \cite{bib12} give the same result with a positive initial constant $A_\oslash=\frac{1}{A^2}$ and use the $\pm$ sign to indicate both backward and inward motion. Instead, our method allows the initial condition $A$ to have positive and negative values. Furthermore, it is directly related to the conserved conjugate momentum $P_\chi=mA$, offering a more straightforward interpretation.
\subsection{Peculiar Velocity, Physical energy and 3- momentum}\label{subsec.3.3}
The peculiar radial velocity of freely-falling particles can be written as
\begin{equation}
v_{\text{pec}}(t) = a(t) \dot{\chi}(t) = \frac{A}{\sqrt{a^2(t) + A^2}}
\label{eq37}.
\end{equation}
Due to the homogeneous and isotropic nature of our FLRW spatial space, this formula applies to all geodesic motions, including non-radial geodesics. Hence, Eq. \eqref{eq37} serves as a general formula for calculating the magnitude of the peculiar velocity of a freely-falling particle. Substituting formula \eqref{eq37} into Eq. \eqref{eq25}, we can derive the expression for the Lorentz factor as
\begin{equation}
\gamma(t) = \frac{dt}{d\tau} = \sqrt{1 + \frac{A^2}{a^2(t)}}
\label{eq38},
\end{equation}
which is consistent with the result in Ref. \cite{bib15}. The 4- momentum vector is now expressed as
\begin{equation}
P^\mu = \left(m\sqrt{1 + \frac{A^2}{a^2(t)}}, \frac{mA}{a^2(t)}, 0, 0\right)
\label{eq39}.
\end{equation}
One can see that $P^\chi = \frac{P_\chi}{a^2(t)}$ as expected from Eqs. \eqref{eq18} and \eqref{eq32}. The physical energy and 3-momentum of the freely-falling particle in FLRW spacetime are given by
\begin{subequations}
\begin{align}
E(t) = P^t = \sqrt{m^2 + \frac{m^2 A^2}{a^2(t)}}
\label{eq40a}\\
p(t) = a(t) P^\chi = \frac{mA}{a(t)}
\label{eq40b},
\end{align}
\end{subequations}
where the energy-momentum relation is well checked.
\subsection{Results and Discussion}\label{subsec.3.4}
From Eq. \eqref{eq37}, the sign of the arbitrary real number $A$ (constant of motion) determines the comoving direction of the free particle. If $A > 0$, it implies that the particle is "comovingly" moving radially outward from the center of the comoving frame ($v_\text{pec}>0$). Conversely, if $A < 0$, the particle is comovingly moving radially inward toward the center of the comoving frame ($v_\text{pec}<0$). In the case where $A = 0$, the particle remains "comovingly" at rest ($v_\text{pec}=0$). However, for $A = \pm \infty$, whene $A \gg a$, the particle will move at the speed of light\footnote{In the next subsection, we will show that this case is related to null-geodesics where $m=0$.}. It is obvious that regardless of the specific value of $A$, we have $0 \leq |v_{\text{pec}}(t)| < 1$; the peculiar velocity of a massive particle could only approach the speed of light but can never reach it. As the universe expands, Eq. \eqref{eq37} shows that the peculiar velocity of a freely-falling particle decreases with time. In the early universe when $a \rightarrow 0$, regardless of the specific value of $A$ (but non-zero), the peculiar velocity of all freely-falling particles tends to the speed of light. In standard cosmology and during the early epochs of the universe, we consider all these particles (which are relativistic) as part of the radiation component of the universe. Through the ages, the scale factor increases, giving rise to a corresponding reduction in the peculiar velocity until it reaches zero when $a \gg A$. In this case, particles can naturally align with the Hubble flow (for limitations on this alignment, see Ref. \cite{bib13}).
\subsection{Null-Geodesic limit}\label{subsec.3.5}
In order to give physical significance to the comoving 4-momentum \eqref{eq39}, both the mass $m$ and the factor $mA$ must be finite. Consequently, when considering $A$ equal to infinity, the mass $m$ must be zero. Accordingly, these two limits are interconnected, both leading to the zero-mass limit. Thus, one can consider null geodesics as the limit when the mass $m$ of particles tends to zero, or equivalently, when the arbitrary constant $A$ approaches infinity
\begin{equation}
A = \pm \infty \iff m = 0
\label{eq41}.
\end{equation}
A more rigorous derivation of geodesics for massless particles from an action principle can be found in Ref. \cite{bib18}. Since we have set the expansion scale factor to be one at the present time, $a(t_0) = 1$, we can infer the physical meaning of the constant of motion $P_\chi$ by evaluating its relation in Eq. \eqref{eq32} at the present time $t_0$ as
\begin{align}
P_\chi= mA = a(t) p(t) = a(t_0) p(t_0) = p_0
\label{eq42},
\end{align}
where $p_0 = p(t_0)$ represents the physical 3-momentum at the present time. Now, we write a suitable formula of the 3-momentum for both massive and massless particles as
\begin{equation}
p(t) = \frac{p_0}{a(t)}
\label{eq43}.
\end{equation}
It is ovious to see that the physical 3-momentum for both massive and massless particle decays as the universe expands.
\subsection{The initial condition for peculiar velocity}\label{subsec.3.6}
From Eqs. \eqref{eq37}, \eqref{eq39}, it becomes apparent that to determine both the peculiar velocity and the comoving 4-momentum of the particle, it is necessary to fix the integration constant $A$ (by correspondingly fixing the constant of motion $P_\chi = p_0$). Instead of relying on $A$ to constrain the motion, a more practical approach involves using a measured quantity such as the initial peculiar velocity. To achieve this, $A$ can be expressed in terms of the initial peculiar velocity $v_{\text{pec}}(t_i)$ at a chosen initial time $t_i$. This can be done by substituting $t = t_i$ into equation \eqref{eq37} and subsequently solving it for $A$. Alternatively, this can be directly derived from the constant of motion formula \eqref{eq32} as
\begin{equation}
A = \frac{P_\chi}{m} = \frac{a(t_i) v_{\text{pec}}(t_i)}{\sqrt{1 - v_{\text{pec}}^2(t_i)}}
\label{eq44}.
\end{equation}
By substituting the expression of $A$ derived earlier into Eqs. \eqref{eq34} and \eqref{eq37}, we obtain
\begin{equation}
\dot{\chi}(t) = \frac{a_i v_i}{a(t) \sqrt{a^2(t)(1 - v_i^2) + a_i^2 v_i^2}}
\label{eq48}.
\end{equation}
and
\begin{equation}
v_{\text{pec}}(t)=a(t) \dot{\chi}(t)= \frac{a_i v_i}{\sqrt{a^2(t)(1-v_i^2) + a_i^2 v_i^2}}
\label{eq45},
\end{equation}
where we have used the notation $a(t_i) = a_i$ and $v_{\text{pec}}(t_i) = v_i$. This relation determines the peculiar velocity of a freely-falling particle at any time $t$ in terms of the known initial peculiar velocity $v_i$ at the initial time $t_i$.
\section{Geodesics Parametrization}\label{sec4}
Geodesic motion is given by integrating over the comoving radial velocity, we have
\begin{equation}
\chi = \chi_i + \int_{t_i}^t \dot{\chi}(t) \, dt
\label{eq51},
\end{equation}
$\chi_i$ represents the initial comoving radial distance at time $t_i$. Following our previous discussion on this topic, we introduce two methods to characterize  a specific geodesic solution of the freely-falling test particle as follows:
\begin{enumerate}[leftmargin=*]
\item $(\chi_i, v_i)$ initial conditions:\\
Using Eqs. \eqref{eq48} and \eqref{eq51}, we write the comoving radial distance $\chi$ of a test-particle for any given time $t$, under two specific initial conditions $(\chi_i, v_i)$ at time $t_i$ as
\begin{equation}
\chi(t;\chi_i, v_i) =\chi_i + \int_{t_i}^t \frac{a_i v_i}{a(t')\sqrt{a^2(t')(1 - v_i^2) + a_i^2 v_i^2} }\, dt'
\label{eq53}.
\end{equation}
We are free to select the initial time $t_i$ to fix the initial comoving distance $\chi_i$ and initial peculiar velocity $v_i$. However, there are some limitations in Eq. \eqref{eq53} when considering the choice $t_i = 0$. In this scenario, $a_i = 0$ and $v_i = \pm 1$ leading to an undefined integral term of $\frac{0}{0}$. This issue arises from the fact that the peculiar velocity for all freely-falling particles at the Big Bang singularity $t = 0$ is equal to the speed of light, making it an unsuitable choice for fixing the initial peculiar velocity $v_i$. To address this problem, one can choose the present time $t_0$ as the reference time to fix the peculiar velocity $v_0 = v_{\text{pec}}(t_0)$, while still considering the Big Bang singularity time $t_i = 0$ for fixing the initial comoving distance $\chi_i = \chi(0)$.
\item $(\chi_i, v_0)$ initial conditions:\\
We can write the comoving radial distance $\chi$ of a test-particle for any given time $t$, under two specific initial conditions $(\chi_i, v_0)$ where $\chi_i$ is the initial comoving distance at an initial time $t_i$ and $v_0$ represents the peculiar velocity at the present time $t_0$, we have
\begin{equation}
\chi(t;\chi_i, v_0) =\chi_i + \int_{t_i}^t \frac{v_0}{a(t')\sqrt{a^2(t')(1 - v_0^2) + v_0^2}} \, dt'
\label{eq52}.
\end{equation}
where the peculiar velocity $v_{\text{pec}}(t)$ at any time $t$, can be expressed in terms of its value $v_{\text{pec}}(t_0) = v_0$ at the present time $t_0 = 0$ simply by setting $i = 0$ in Eq. \eqref{eq45}, which gives
\begin{equation}
v_{\text{pec}}(t) = \frac{v_0}{\sqrt{a^2(t)(1 - v_0^2) + v_0^2}}
\label{eq46},
\end{equation}
where we have used $a_0 = 1$. $v_0$ can be considered as the measured physical velocity of the free particle from a nearby comoving observer (to neglect Hubble flow) at the present time. For a distant object, we should take into account the Hubble flow effect, and we must subtract the recessional velocity to get the actual peculiar velocity. We shall use this parametrization \eqref{eq52} to study the behavior of geodesics in terms of an initial distance $\chi_i$ at the Big Bang singularity $t_i=0$.
\end{enumerate}
Since the comoving radial distance $\chi$ is inherently positive, the right-hand side of Eqs. \eqref{eq51}, \eqref{eq53}, and \eqref{eq52} should be within an absolute value. However, the absolute value is removed to allow the comoving radial distance to be either positive or negative. This distinction is crucial for indicating the direction of motion, i.e., $\chi>0$ for the starting motion direction and $\chi<0$ for the opposite one, especially when a free particle crosses our position from the positive to the opposite direction.
\begin{itemize}[leftmargin=*]
\item \textbf{Comoving geodesic limit}\\
if we set $v_0 = \pm 1$ in Eq. \eqref{eq52} (or $v_i = \pm 1$ in Eq. \eqref{eq53}), we obtain the corresponding null geodesic
\begin{equation}
\chi(t; \chi_i, \pm 1) = \chi_i \pm \int_{t_i}^t \frac{dt'}{a(t')}
\label{eq54},
\end{equation}
\item \textbf{Null geodesic limit}\\
If we take $v_0 = 0$ in Eq. \eqref{eq52} (or $v_i = 0$ in Eq. \eqref{eq53}), the particle remains at a constant comoving distance
\begin{equation}
\chi(t;\chi_i, 0) = \chi_i
\label{eq55}.
\end{equation}
\end{itemize}
In our next analysis, we will use the geodesic relation detailed in Eq. \eqref{eq52}, using the conditions $(\chi_i, v_0)$. This choice is significantly better for discussing a geodesic motion starting from the Big Bang singularity $t_i=0$, which serves as our next illustrative example. However, Eq. \eqref{eq53} is more useful for practical observations such as those involving a galaxy with initial time $t_i \neq 0$.
\section{Application: $\Lambda$CDM Model}\label{sec5}
\subsection{$\Lambda$CDM Model}\label{subsec.5.1}
A cosmological model is a scientific framework for comprehending the universe, including its evolution, structure, and composition. Cosmological models are based on observations of the universe, such as the distribution of galaxies and the CMB. Through comparison with actual observations, the accuracy of the models is assessed, leading to continual refinement. The most widely accepted cosmological model is the $\Lambda$CDM model. The is model states that the universe is composed of basic components with dimensionless density at present time of about (according to Planck Collaboration 2018 \cite{bib3}):
\begin{align*}
\text{Radiation:} & \quad \Omega_r \approx 9 \times 10^{-5} \\
\text{Matter:} & \quad \Omega_m \approx 0.315 \\
\text{Cosmological constant:} & \quad \Omega_\Lambda \approx 0.685
\end{align*}
and it is based on the Friedmann-Lemaître equation
\begin{equation}
  \frac{\dot{a}(t)}{a(t)} = H(t) = H_0 \sqrt{\frac{\Omega_r}{a^4(t)} + \frac{\Omega_m}{a^3(t)} + \frac{\Omega_k}{a^2(t)} + \Omega_\Lambda}
  \label{eq56}.
\end{equation}
$H_0$ is the Hubble parameter today (according to Plank 2018): $H_0 \approx 67.4 \, \left(\frac{\text{km} \cdot \text{s}^{-1}}{\text{Mpc}}\right)$.
We can write the Friedmann-Lemaître equation in the simple form
\begin{equation}
H(t)=H_0 E(t)
\label{eq57},
\end{equation}
where the dimensionless Hubble parameter
\begin{equation}
E(t)=\sqrt{\frac{\Omega_r}{a^4(t)} + \frac{\Omega_m}{a^3(t)} + \frac{\Omega_k}{a^2(t)} + \Omega_\Lambda}
\label{eq58}.
\end{equation}
If we neglect the radiation and curvature density parameters $\Omega_r \approx \Omega_k \approx0$, it becomes straightforward to analytically solve the Friedman-Lemaître equation \eqref{eq56} for the scale factor $a(t)$ as follows
\begin{equation}
a(t) = \left(\frac{\Omega_m}{\Omega_\Lambda}\right)^{1/3} \sinh^{2/3}\left(\frac{3}{2} \sqrt{\Omega_\Lambda}H_0t\right)
\label{eq59}.
\end{equation}
We will discuss in what follows three cosmological concepts:
\begin{itemize}[leftmargin=*]
\item \textbf{The age of the universe:} It is the present time $t_0$ and it can be calculated as:
\begin{equation}
t_0 = \int_0^{t_0} dt = \frac{1}{H_0} \int_0^1 \frac{da}{aE(a)} \approx 13.79 \, \text{Gyr}
\label{eq60}
\end{equation}
\item \textbf{The Particle Horizon at $t=t_0$:} It is the comoving\footnote{It is also the physical radius of the current observed universe, since we have considered $a(t_0) = 1$.} distance $\chi_0$ from the Earth to the current edge of the observable universe
\begin{equation}
\chi_0 = \int_0^{t_0} \frac{dt'}{a(t')} = \frac{1}{H_0} \int_0^1 \frac{da}{a^2 E(a)} \approx 47 \, \text{Gly}
\label{eq61}
\end{equation}
\item \textbf{The Particle Horizon at $t\rightarrow +\infty$:} It is the comoving distance $\chi_\infty$ from Earth to the edge of the observable universe when time goes to infinity\footnote{It is also the event horizon at $t=0$.}
\begin{equation}
\chi_\infty = \int_0^{+\infty} \frac{dt'}{a(t')} = \frac{1}{H_0} \int_0^{+\infty} \frac{da}{a^2 E(a)} \approx 63.68 \, \text{Gly}
\label{eq62}
\end{equation}
\end{itemize}
Now we will apply these results to the $\Lambda$CDM model by drawing some illustrative graphs. We will focus our analysis in the comoving coordinate system, and every time we need to revert the physical picture, one simply multiply the comoving distance by the scale factor $a(t)$.
\subsection{Geodesics, Past Light cone, and Horizons}\label{subsec.5.2}
In this section, we demonstrate how the geodesic equation of motion $\chi(t;\chi_i, v_0)$ in Eq. \eqref{eq52} provides the expressions of the well-known cosmological horizons and past light cone. For this study, we fix the initial time $t_i$ of geodesic motion at the moment of the Big Bang singularity $t_i = 0$. Our interest primarily lies in geodesics originating from $t=0$. This is because any geodesic characterized by $(\chi_i, v_0)$ that begins moving at $t_i>0$ can be traced backward in time and matched with a geodesic of $t_i=0$ under different initial conditions $(\chi'_i, v_0)$, while maintaining the same peculiar velocity $v_0$. In what follows, we adopt the units of distance and time in billions of light-years (Gly) and billions of years (Gyr), respectively, so that $H_0^{-1} \approx 14.51$ Gyr. We position ourselves as a comoving observer at the origin $\chi = 0$ disregarding any peculiar velocities attributable to the Milky Way Galaxy, the Solar System, or Earth's own motion. We shall compute the integrals and generate the corresponding curves numerically.
\begin{itemize}[leftmargin=*]
\item \textbf{The past light cone:}
It is the geodesic motion for an incoming light beam, characterized by a peculiar velocity $v_0 = -1$ at the present time, starting from an initial comoving radial distance $\chi_i = \chi_0$ \eqref{eq61} at $t=0$
\begin{equation}
\chi(t;\chi_0,-1) = \int_0^{t_0} \frac{dt'}{a(t')} - \int_0^t \frac{dt'}{a(t')} = \int_t^{t_0} \frac{dt'}{a(t')}
\label{eq63}
\end{equation}
\item \textbf{The event Horizon:}
It is the geodesic motion for an incoming light beam, characterized by a peculiar velocity $v_0=-1$ at the present time, starting from an initial comoving radial distance $\chi_i = \chi_\infty$ \eqref{eq62} at $t=0$
\begin{equation}
\chi(t;\chi_\infty,-1) = \int_0^{+\infty} \frac{dt'}{a(t')} - \int_0^t \frac{dt'}{a(t')} = \int_t^{\infty} \frac{dt'}{a(t')}
\label{eq64}
\end{equation}
\item \textbf{The particle Horizon:}
It is geodesic motion for an outgoing light beam $v_0 = +1$ at the present time, starting from our location $\chi_i = 0$ at $t=0$
\begin{equation}
\chi(t; 0, +1) = 0 + \int_{0}^{t} \frac{dt'}{a(t')} = \int_{0}^{t} \frac{dt'}{a(t')}
\label{eq65}
\end{equation}
\item \textbf{Generic Geodesic Motion:}
We will consider an example of a test-particle starting from an initial comoving distance $\chi_i = 50$ Gly at $t=0$ possessing a range of selected peculiar velocities:
\begin{equation}
v_0 = \{0, \pm0.2, \pm0.4, \pm0.6, \pm0.8, \pm1\}
\label{eq66},
\end{equation}
at the present time. One can plot graphs for all selected peculiar velocities in the comoving distance framework. Additionally, it is possible to add the Hubble sphere, defined by the comoving radius $\chi_H(t) = (a(t)H(t))^{-1}$; this radius determines the distance at which the recession velocity of an object, due to the expansion of the universe, is equal to the speed of light. To obtain analogous graphs within the physical distance framework, one simply multiplies the corresponding formulas by $a(t)$. For example, the radial physical distance $\chi_{\text{phys}}$ is given by
\begin{equation}
\chi_{\text{phys}}(t; \chi_i, v_0) = a(t) \chi(t; \chi_i, v_0)
\label{eq67}.
\end{equation}
Upon taking its time derivative, one can check that the formula for the physical velocity, as presented in Eq. \eqref{eq3b}, is well satisfied.
\end{itemize}
\begin{figure}[htbp]%
\centering
\includegraphics[width=0.8\textwidth]{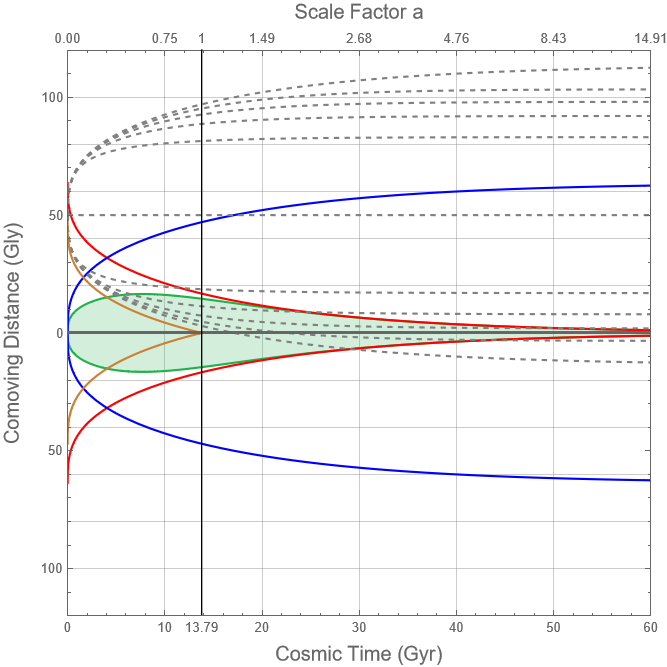}
\caption{Comoving distance frame: The black solid lines indicate: (horizontal) our comoving worldline, (vertical) our present universe. The blue line is the particle horizon, the red line is the event horizon, the green area is the Hubble sphere. The dashed grey curves are the free-falling geodesics corresponding to a test-particle with an initial comoving distance of $50$ Gly at $t=0$ and for radial peculiar velocities at the present time $-1, -0.8, -0.6, -0.4, -0.2, 0, 0.2, 0.4, 0.6, 0.8, 1$ from the bottom curve to the top one, respectively. The horizontal gridlines correspond to comoving worldlines. One can see that all freely-falling particles will become almost comoving when time goes to infinity.
}\label{fig3}
\end{figure}
\begin{figure}[htbp]%
\centering
\includegraphics[width=0.8\textwidth]{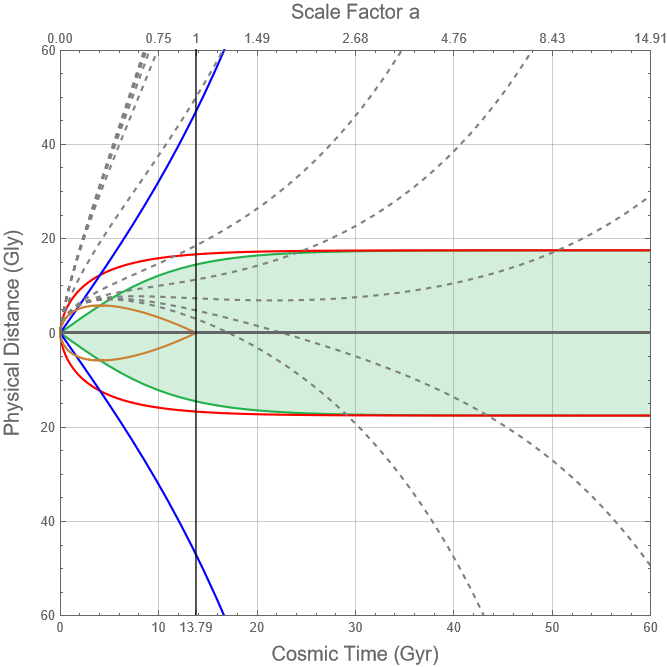}
\caption{Physical distance frame: The black solid lines indicate: (horizontal) our comoving worldline, (vertical) our present universe. The orange line is the past light cone, the green area is the Hubble sphere, the blue line is the particle horizon, the red line is the event horizon. The dashed grey curves are the free-falling geodesics corresponding to a freely-falling test-particle with a chosen initial comoving distance of $50$ Gly at $t=0$ and for radial peculiar velocities at the present time: $-1, -0.8, -0.6, -0.4, -0.2, 0, 0.2, 0.4, 0.6, 0.8, 1$ from the bottom curve to the top one, respectively. This graph shows the actual physical distance-time diagram for freely-falling particles in FLRW spacetime.
}\label{fig4}
\end{figure}
\section{Physical Distance, Velocity and Acceleration}\label{sec6}
To rigorously discuss the physical motion $\chi_{\text{phys}}$ of the freely-falling particle in FLRW spacetime, we need to determine the roots of the first derivative $\dot{\chi}_{\text{phys}}$ (physical velocity), as well as the sign of the corresponding second derivative $\ddot{\chi}_{\text{phys}}$ (physical acceleration). Our initial step involves calculating both $\dot{\chi}_{\text{phys}}$ and $\ddot{\chi}_{\text{phys}}$. We begin with the equation
\begin{align}
\chi_{\text{phys}}(t)&= a(t)\chi(t)\nonumber\\
&= a(t)\left[\chi_i + \int_{t_i}^{t} \frac{v_0}{a(t')\sqrt{a^2(t')(1 - v_0^2) + v_0^2}} dt'\right]
\label{eq68}
\end{align}
Taking its time derivative yields the following expression
\begin{align}
\dot{\chi}_{\text{phys}}(t) &= \dot{a}(t)\chi(t) + a(t)\dot{\chi}(t)\nonumber\\
&= H(t)\chi_{\text{phys}}(t) + v_{\text{pec}}(t)
\label{eq69}
\end{align}
As expected, we arrive at the well-established formula for physical velocity that has been discussed in Eq. \eqref{eq3b}. Applying the time derivative once more, we obtain
\begin{align}
\ddot{\chi}_{\text{phys}}(t) &= \dot{v}_{\text{rec}}(t) + \dot{v}_{\text{pec}}(t)\nonumber\\
&= \frac{\ddot{a}(t)}{a(t)}\chi_{\text{phys}}(t) + \frac{\dot{a}(t)}{a(t)}v_{\text{pec}}^3(t)
\label{eq70}
\end{align}
where separate calculations of the recessional and peculiar accelerations yield,
\begin{itemize}[leftmargin=*]
\item \textbf{For recessional acceleration $\gamma_{\text{rec}}$:}
\begin{align}
\gamma_{\text{rec}}(t) &= \dot{v}_{\text{rec}}(t)\nonumber\\
&= \frac{\ddot{a}(t)}{a(t)}\chi_{\text{phys}}(t) + \frac{\dot{a}(t)}{a(t)}v_{\text{pec}}(t)
\label{eq71}
\end{align}
\item \textbf{For peculiar acceleration $\gamma_{\text{pec}}$:}
\begin{align}
\gamma_{\text{pec}}(t)& = \dot{v}_{\text{pec}}(t)\nonumber\\
&= \frac{\dot{a}(t)}{a(t)}v_{\text{pec}}^3(t) - \frac{\dot{a}(t)}{a(t)}v_{\text{pec}}(t)
\label{eq72}
\end{align}
\end{itemize}
In the derivation of $\dot{v}_{\text{pec}}$, we have benefited from the fact that $A$ in Eq. \eqref{eq44} is a conserved quantity throughout the motion. The formula \eqref{eq70} is the same result found in Ref. \cite{bib12}. We derive the formula \eqref{eq70} for the physical acceleration of a freely-falling particle in FLRW spacetime, which can be decomposed into two components as follows
\begin{equation}
\gamma_{\text{phy}}(t) = \ddot{\chi}_{\text{phys}}(t) = \gamma_{r}(t) + \gamma_{p}(t)
\label{eq73}.
\end{equation}
\begin{itemize}[leftmargin=*]
\item \textbf{Quasi-recessional acceleration $\gamma_r$:}
\begin{equation}
\gamma_{r}(t) = \frac{\ddot{a}(t)}{a(t)}\chi_{\text{phys}}(t) = \ddot{a}(t)\chi(t)
\label{eq74}
\end{equation}
\item \textbf{Quasi-peculiar acceleration $\gamma_p$:}
\begin{equation}
\gamma_{p}(t) = \frac{\dot{a}(t)}{a(t)}v_{\text{pec}}^3(t) = \frac{\dot{a}(t)}{a(t)}\frac{v_0^3}{(a^2(t)(1 - v_0^2) + v_0^2)^{3/2}}
\label{eq75}
\end{equation}
\end{itemize}
We use the term “quasi” to indicate that $\gamma_r$ and $\gamma_p$ are not really the recessional $\gamma_{\text{rec}}$ and the peculiar $\gamma_{\text{pec}}$ acceleration, respectively.
\section{Geodesic Behavior in $\Lambda$CDM Model}\label{sec7}
With these results in hand, we are able to investigate the behavior of a freely-falling particle in FLRW spacetime.
\subsection{At the BB singularity $t=0$}\label{subsec.7.1}
Before delving into the analysis, it is important to recognize from Figs. ~\ref{fig3} and ~\ref{fig4} that all geodesics start moving away from us, even if the particle exhibits peculiar motion in our direction characterized by a negative radial peculiar velocity. This observation can be proved through equations \eqref{eq37} and \eqref{eq69}, which demonstrate that the physical radial velocity of any test particle at  $t = 0$ will be positively infinite
\begin{equation}
\dot{\chi}_{\text{phys}}(0) = \dot{a}(0) \chi_i + \text{sign}(v_0) \rightarrow +\infty
\label{eq76},
\end{equation}
where $\dot{a}(0)$ goes to infinity. The origin of this behavior lies in the fact that the recessional velocity starts with an infinitely positive value at $t=0$, while the peculiar velocity remains perpetually constrained below the speed of light. Consequently, it can be inferred that the Hubble flow dominates in the initial stages of all geodesic motions.
\subsection{After the BB singularity $t>0$}\label{subsec.7.2}
As the universe expands, the cosmic expansion rate slows down due to the gravitational attraction between all forms of matter and radiation which act against the expansion, causing the recessional velocity to decrease over time. For geodesics with a positive peculiar velocity (particles in peculiar motion away from us $v_{\text{pec}} > 0$), the particles will continue moving away from us (one-way journey). Conversely, for geodesics with a negative peculiar velocity (particles in peculiar motion toward us, $v_{\text{pec}} < 0$) that will be the primary focus of our next study; while some will continue to recede (one-way journey), others possess a sufficiently large negative peculiar velocity that ultimately balances the decreasing recessional velocity. This equilibrium occurs when the recessional velocity decreases to a point where it equals the magnitude of the negative peculiar velocity, leading to their mutual cancellation. At this equilibrium time, denoted as $t_*$, the particle attains zero physical velocity, when the radial physical distance is extremized
\begin{equation}
\dot{\chi}_{\text{phys}}(t_*,\chi_i, v_0) = 0
\label{eq77},
\end{equation}
which can be proved from Eq. \eqref{eq69} that corresponds to
\begin{equation}
\chi_i = f(t_*, v_0)
\label{eq78},
\end{equation}
where
\begin{equation}
f(t_*, v_0) = -\frac{1}{\dot{a}(t_*)} \frac{v_0}{\sqrt{a^2(t_*)(1 - v_0^2) + v_0^2}}
- \int_{0}^{t_*} \frac{v_0}{a(t')\sqrt{a^2(t')(1 - v_0^2) + v_0^2}} dt'
\label{eq79}
\end{equation}
Furthermore, the sign of the second derivative $\ddot{\chi}_{\text{phys}}$ of the physical coordinate must be used to ensure whether the observed extremum corresponds to a maximum, minimum, or an inflection point. One can show that
\begin{equation}
\ddot{\chi}_{\text{phys}}(t_*, \chi_i, v_0) = 0
\label{eq80},
\end{equation}
corresponds to the following relation
\begin{equation}
\chi_i = g(t_*, v_0)
\label{eq81},
\end{equation}
where
\begin{equation}
g(t_*, v_0) = -\frac{\dot{a}(t_*)}{\ddot{a}(t_*)} \frac{v_0^3}{[a^2(t_*)(1 - v_0^2) + v_0^2]^{3/2}}
- \int_{0}^{t_*} \frac{v_0}{a(t')\sqrt{a^2(t')(1 - v_0^2) + v_0^2}} dt'
\label{eq82}
\end{equation}
Now, the condition for these extrema to be maximum, minimum, or an inflection point is shown to be related to the sign of $\ddot{a}(t_*)$ and the two functions $f$ and $g$ in Eqs. \eqref{eq79} and \eqref{eq82} as delineated in Table. ~\ref{tab1}.
\begin{table*}[t]
\centering
\begin{NiceTabular}{|c|c|c|c|c|c|c|}[cell-space-limits=5pt]
\Xhline{2pt}
&$\gamma_r (t_*)$&$\gamma_p (t_*)$&$\ddot{\chi}_{\text{phys}}(t_*)$&condition (1)&condition (2)&\makecell{kinds\\of extrema}\\ \Xhline{1.7pt}
$\ddot{a}(t_*) < 0$&$-$&$-$&$-$& \Block{4-1}{\makecell{$\dot{\chi}_{\text{phys}}(t_*) = 0$ \\ $\Updownarrow$\\$\chi_i = f(t_*, v_0)$}} &$0 < t_* < 7.69 \, \text{By}$& maximum\\
\cmidrule{1-4} \cmidrule{6-7} 
$\ddot{a}(t_*) = 0$&$0$&$-$&$-$&&$t_* = 7.69 \, \text{By}$&maximum\\
\cmidrule{1-4} \cmidrule{6-7} 
\Block{3-1}{\makecell{$\ddot{a}(t_*) > 0$\\$t_{**} \geq t_*$}} & \Block{3-1}{$+$} & \Block{3-1}{$-$} &$-$&&\makecell{$\chi_i < g(t_*, v_0)$\\$t_* > 7.69 \, \text{By}$}&maximum\\ \cmidrule{4-4} \cmidrule{6-7} 
&&&$0$ &&\makecell{$\chi_i = g(t_*, v_0)$\\$t_* > 7.69 \, \text{By}$}&\makecell{inflection\\point}\\ \cmidrule{4-7} 
&&&$+$&\makecell{$\dot{\chi}_{\text{phys}}(t_{**}) = 0$ \\ $\Updownarrow$\\$\chi_i = f(t_{**}, v_0)$}&\makecell{$\chi_i > g(t_{**}, v_0)$\\$t_{**} > 7.69 \, \text{By}$}&minimum\\ \Xhline{2pt}
\end{NiceTabular}
\caption{Summary of conditions for different types of extrema of the physical radial distance $\chi_\text{phys}$.}
\label{tab1}
\end{table*}
\\

\begin{figure}[t]%
\centering
\includegraphics[width=0.6\textwidth]{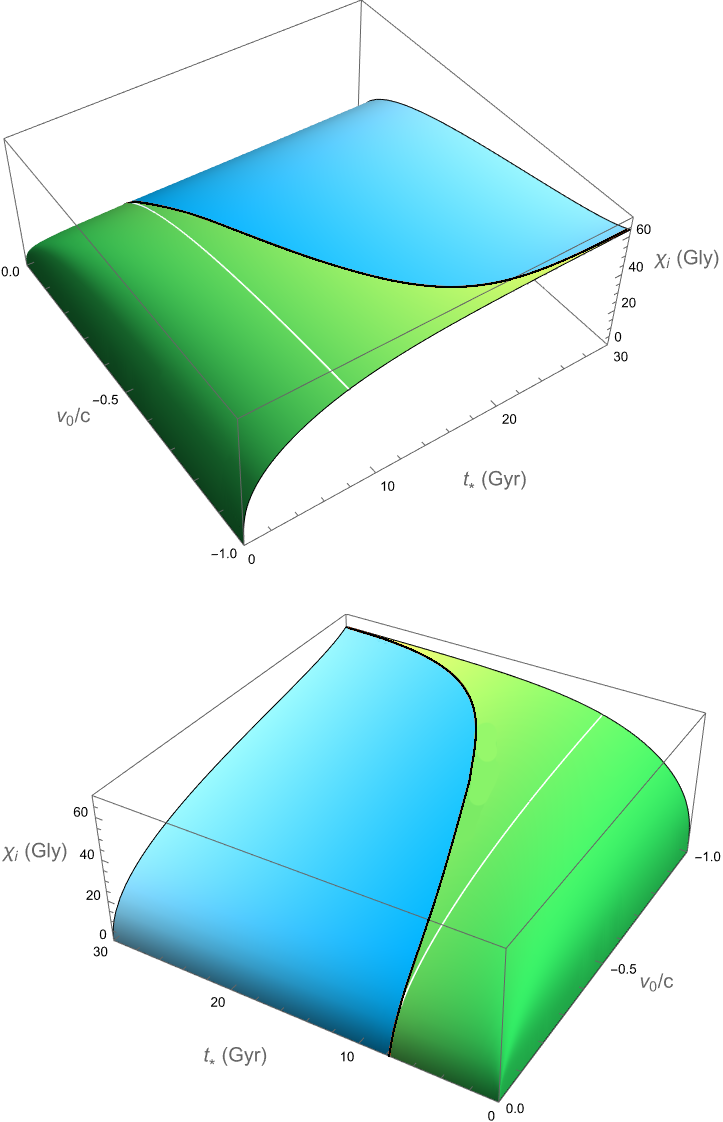}
\caption{It illustrates the temporal condition $t_*$ under which a free-falling particle with an initial condition $(\chi_i,v_0)$ experiences a maximum or minimum in its physical radial distance, $\dot{\chi}_{\text{phys}} = 0$. The green surface denotes a local maximum, whereas the blue surface indicates a local minimum. The black curve between these surfaces corresponds to the inflection points. The white line represents the constant time slice $t=7.69  \ \text{Gyr}$ when the expansion of the universe began to accelerate.}\label{fig5}
\end{figure}

All the outcomes listed in Table. ~\ref{tab1} are readily proved with the use of Eqs. \eqref{eq79} and \eqref{eq82}. For instance, the quasi-peculiar acceleration $\gamma_p(t_*)$ given in Eq. \eqref{eq75} is always negative since we are focusing on the case of $v_{\text{pec}} < 0$. The sign of the quasi-recessional acceleration $\gamma_r(t_*)$, on the other hand, aligns with that of $\ddot{a}(t_*)$, as indicated in Eq. \eqref{eq74}. In the context of our cosmological model, $\ddot{a}(t)$ becomes zero at $t = 7.69 \ \text{Gyr}$, signaling the transition to an accelerating expansion of the universe. In our analysis, we seek to identify the geodesics characterized by $(\chi_i, v_0)$ which result in the vanishing of the physical velocity at time $t_*$; it is achieved through the construction of a 3D surface plot that represents the conditions in Table. ~\ref{tab1} for extremizing the physical radial distance \eqref{eq67}, as depicted in Fig. ~\ref{fig5}.
\begin{figure}[htbp]%
\centering
\includegraphics[width=0.7\textwidth]{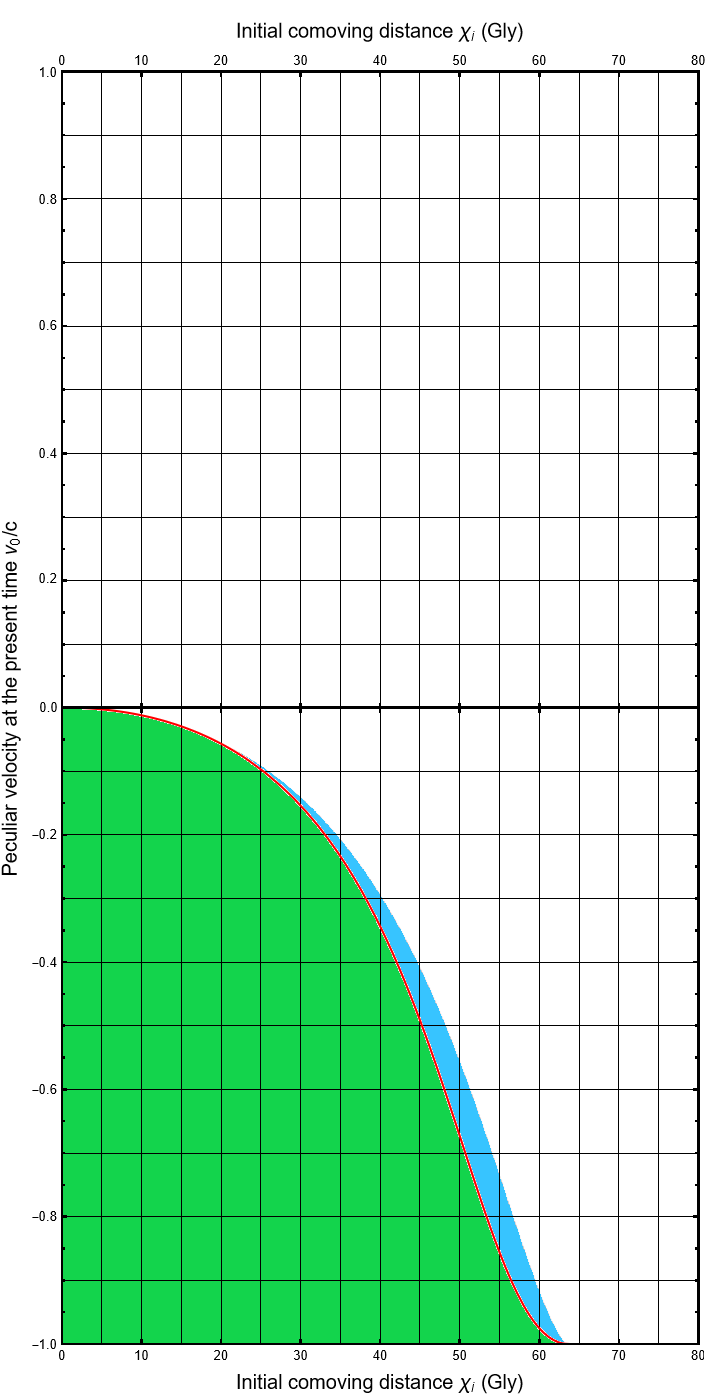}
\caption{It is the projection of the 3d- Fig. ~\ref{fig5} onto a 2d- $(\chi_i, v_0)$ plane where the blue region dotted with green corresponds to the overlapping projection of the blue and green surfaces in the 3d- Fig. ~\ref{fig5}. It illustrates the initial conditions $(\chi_i, v_0)$ for various geodesics starting at the initial time $t=0$. Each point $(\chi_i, v_0)$ represents a distinct geodesic path for a free particle in FLRW spacetime. The figure includes both negative and positive peculiar velocities. (0) the uncolored area denotes geodesics of one-way journey perpetually receding from us, while all colored areas indicate return journey geodesics: (1) the blue area dotted with green represents the return geodesics that will eventually move away from us, (2) the green area for those crossing and surpassing our position, and (3) the red line for return geodesics approaching us without crossing. The boundary curve where the blue and uncolored areas intersect signifies also a one-way journey, characterized by a momentary pause at the inflection point.
}\label{fig6}
\end{figure}
\subsection{After the  equilibrium time $t>t_*$}\label{subsec.7.2}
Since the expressions for recessional velocity and peculiar velocity are fundamentally different. As a result, after the time $t_*$, one of them is bound to dominate the other, indicating that the equilibrium at $t_*$ is unstable. Therefore, it is impossible for the freely-falling particle at non-vanishing radial distance to remain at rest relative to our physical reference frame. Accordingly, after crossing the temporal point $t_*$, the behavior of particles is determined by the dominant velocity component:
\begin{itemize}[leftmargin=*]
\item \textbf{One-way journey geodesics:} If the recessional velocity dominates after the time $t_*$, the particles will return to an irreversible outward motion, pursuing their journey outward indefinitely. This case corresponds to the triplet $(t_*, \chi_i, v_0)$ in the black boundary curve between the green and blue surfaces in Fig. ~\ref{fig5}, as well as it is related to the conditions $(\chi_i, v_0)$ belonging to the boundary curve between the blue and uncolored regions in Fig. ~\ref{fig6}. Mathematically speaking, it is the case of the inflection point which satisfies both conditions $\dot{\chi}_{\text{phys}} = 0$ and $\ddot{\chi}_{\text{phys}} = 0$, as presented in the fifth row of Table. ~\ref{tab1}. 
\item \textbf{Return journey geodesics:} For other geodesics, where the peculiar velocity is dominant, there will be a directional change of physical motion, and consequently, the particle will begin moving toward us. This corresponds to the triplet $(t_*, \chi_i, v_0)$ belonging to the green surface in Fig. ~\ref{fig5}, as well as it is related to the conditions $(\chi_i, v_0)$ belonging to colored regions (green and blue) in Fig. ~\ref{fig6}. Mathematically speaking, this case indicates a local maximization of the physical radial distance, which satisfy both conditions $\dot{\chi}_{\text{phys}} = 0$ and $\ddot{\chi}_{\text{phys}} < 0$ at time $t_*$, as presented in the second, third and fourth rows in Table. ~\ref{tab1}.
\end{itemize}
\subsection{Return journey after $t_*$}\label{subsec.7.4}
\begin{figure}[t]%
\centering
\includegraphics[width=0.8\textwidth]{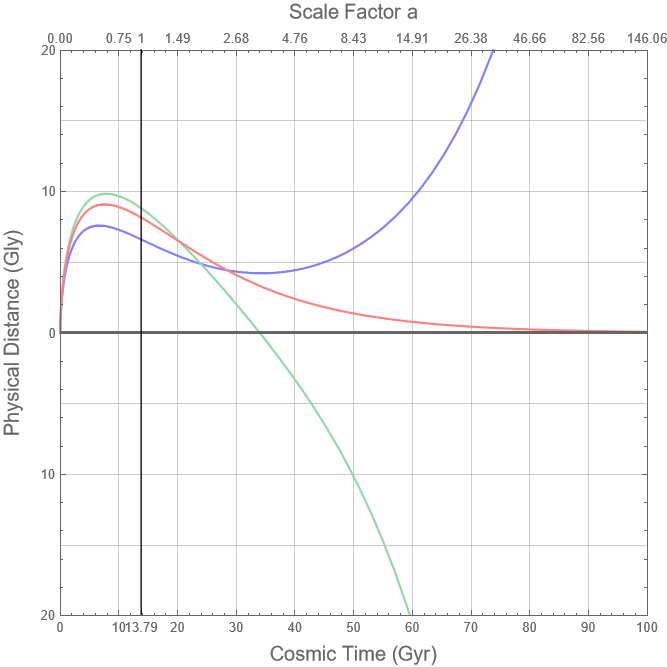}
\caption{Showcases of three distinct return geodesic scenarios in the physical distance frame: 
(1) Blue curve: a return geodesic that resumes moving away from our reference frame, characterized by initial conditions $(\chi_i, v_0) = (50 \text{ Gly}, -0.65c)$. 
(2) Green curve: a return geodesic that crosses our location, identified by initial conditions $(\chi_i, v_0) = (55 \text{ Gly}, -0.9c)$. 
(3) Red curve: a return geodesic converging towards our location, specified by initial conditions $(\chi_i, v_0) \approx (53.37 \text{ Gly}, -0.8c)$.
}\label{fig7}
\end{figure}
During this phase, there are three distinct possibilities for free particles with dominant peculiar velocity moving toward us:
\begin{enumerate}[leftmargin=*]
\item The first possibility implies a temporary dominance of peculiar velocity that may eventually be superseded by the recessional velocity due to the accelerated expansion of the universe, potentially reversing the particle's direction once more. This phenomenon can only occur after the universe enters a phase of accelerated expansion ($t > 7.69$ Gyr). This corresponds to the triplets $(t_{**}, \chi_i, v_0)$ of the blue surface in Fig. ~\ref{fig5}, as well as it is related to the conditions $(\chi_i, v_0)$ belonging to the blue region dotted with green in Fig. ~\ref{fig6}. Mathematically speaking, this occurs when the physical radial distance extremized again and making a local minimum at some time $t_{**}$, where $t_{**} > t_*$, which satisfies both conditions $\dot{\chi}_{\text{phys}} = 0$ and $\ddot{\chi}_{\text{phys}} > 0$ at time $t_{**}$ as presented in the last row in Table. ~\ref{tab1}. In the scenario where a returning particle starts to move again away from us, we assume the initial conditions of $(\chi_i,v_0) = (50 \ \text{Gly},-0.65c)$. This specific geodesic is depicted by the blue curve in Fig. ~\ref{fig7}.
\item The second scenario is when the peculiar velocity of these free particles is sufficient to overcome the recessional velocity over an extended period of time, ensuring their continuous motion toward us. Even after the universe enters a phase of accelerated expansion, The particles whose velocity are still dominated by the peculiar component will maintain their inward journey, ultimately surpassing our position. After crossing us, their peculiar velocity will align with the Hubble flow direction, and become positive. Finally, these particles will perpetually recede from us, continuing their outward trajectory on the opposite side indefinitely. This corresponds to the conditions $(\chi_i, v_0)$ belonging to the green region in Fig. ~\ref{fig6}. In that case where the particle is crossing our position, we consider the initial conditions $(\chi_i,v_0) = (55 \ \text{Gly},-0.9c)$. It is depicted by the green curve in Fig. ~\ref{fig7}.
\item The final scenario presents a strange case between the first and second possibilities: free particles will approach us but never actually cross our position. Instead, they continue their approach ad infinitum. As they draw nearer, their peculiar velocity incrementally diminishes until it asymptotically approaches the recessional velocity at infinity, which leads to their cancellation. This Leads to a scenario where the particles maintain a relatively stable position with respect to our reference frame. This corresponds to the conditions $(\chi_i, v_0)$ belonging to the red curve in Fig. ~\ref{fig6}. In this case where the particle is perpetually approaching us, we consider the initial conditions $(\chi_i,v_0) \approx (53.37 \ \text{Gly},-0.8c)$. We get this convergent geodesic represented by the red curve as depicted in Fig. ~\ref{fig7}. The approximation sign "$\approx$" indicates that any slight deviation from the exact  conditions represented by the red curve in Fig. ~\ref{fig6} results in a shift from this scenario to either the first or second scenario. Thus, this possibility is fundamentally unstable, making its precise realization a borderline case.
\end{enumerate}
In the illustrated scenarios of Fig. ~\ref{fig6}, all colored regions represent return journey geodesics. However, only those within the green zone will ultimately reach our location. Consequently, any free particle starting its journey at $t = 0$ with initial conditions outside the green area will not reach our location. Therefore, the red curve delineates the ultimate future horizon, signaling the boundary of the farthest visitors we can expect. Notably, for a free particle moving at the speed of light $v_0 = -1$, its ultimate future horizon is shown to originate from a comoving distance of $\chi_i = \chi_{\infty} \approx 63.68$ Gly, in agreement with the prediction made in Eq. \eqref{eq62}. Importantly, the manifestation of these scenarios is independent of the choice of the initial time $t_i$ in our analysis; changing $t_i$ from zero affects only the initial conditions $(\chi_i, v_0)$ necessary for the realization of each scenario.
\subsection{The asymptotical limit}\label{subsec.7.5}
One can see from Fig. ~\ref{fig3} that all freely-falling particles will eventually join into the Hubble flow as time approaches infinity. This phenomenon is not a universal characteristic of all eternally expanding universes (for more details, see Ref. \cite{bib13}). It specifically occurs under the crucial condition $w_d < -1/3$, where $w_d$ is the equation of state parameter for the dominant cosmic component as time goes to infinity. In our example, the dominant component as $t \rightarrow +\infty$ is the cosmological constant $\Lambda$, characterized by $w_\Lambda = -1$, thus fulfilling the condition for asymptotic rejoining the Hubble flow.
\subsection{Access conditions}\label{subsec.7.6}
\begin{figure}[t]%
\centering
\includegraphics[width=0.8\textwidth]{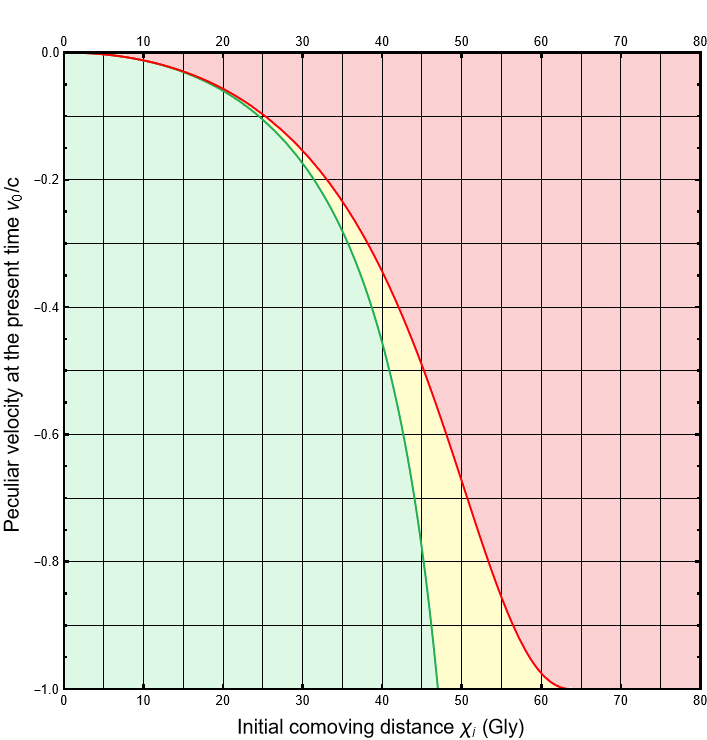}
\caption{It illustrates the initial condition $(\chi_i, v_0)$ for geodesics of radially incoming, freely-falling particles in FLRW spacetime starting from $t=0$. The red curve corresponds to the set of geodesics that asymptotically approach us as time tends toward infinity (our latest visitors). The green curve signifies the set of geodesics whose particles are reaching our position at the present time (our current visitors). The green shaded area maps all geodesics of all free particles that have actually reached us in the past. The yellow area designates geodesics for particles that will reach in the future. Contrarily, the red region denotes the geodesics of particles that will never reach us.
}\label{fig8}
\end{figure}
A freely-falling particle characterized by the initial conditions ($\chi_i,v_0$), will reach our location (the origin $\chi = 0$) at an arbitrary time $t$, if and only if both $\chi_i$ and $v_0$ fulfill the condition
\begin{equation}
\chi(t;\chi_i, v_0) = 0
\label{eq83},
\end{equation}
where the comoving radial distance $\chi$ is given in Eq. \eqref{eq52}. The "access condition" governs whether the particle will reach our location and is expressed as
\begin{equation}
\chi_i = -\int_{0}^{t} \frac{v_0}{a(t') \sqrt{a^2(t')(1 - v_0^2) + v_0^2}} dt'
\label{eq84}.
\end{equation}
The set of geodesics defined by the initial conditions ($\chi_i, v_0$) that satisfy the access condition \eqref{eq84} for $t \rightarrow +\infty$ determines our latest visitors – those freely-falling particles that will eventually reach us as time goes to infinity. Particles with initial conditions beyond this set will never reach our location. The geodesics characterized by initial condition ($\chi_i, v_0$) that satisfy the “access condition” at the present time $t=t_0$ are identified as our current visitors. The possible geodesics for both the current and latest visitors are depicted in Fig. ~\ref{fig8} by the green and red curves, respectively. Notably, the green curve represents the limit of all our observable horizons, including not just incoming light (where $v_0 = -1$), but also any negative peculiar velocity $-1 \leq v_0 < 0$. For $v_0=-1$, the current and latest visors are corresponding to initial comoving distances of $\chi_i = \chi_0 \approx 47 \ \text{Gly}$ and $\chi_i = \chi_{\infty} \approx 63.68 \ \text{Gly}$, in agreement with the predictions made in Eqs. \eqref{eq61} and \eqref{eq62}, respectively.
\subsection{$\dot{\chi}_{\text{phys}}(+\infty) = 0 \iff \chi_{\text{phys}}(+\infty)=0 $}\label{subsec.7.7}
Now, we can notice that the geodesics for the $3^\text{rd}$ scenario of the Return Journey (RJ) represented by the red curve in Fig.  ~\ref{fig6}, precisely match the geodesics for the access condition for our latest visitors shown by the the red curve in Fig.  ~\ref{fig8}. They are corresponding to the same case where free particles return at time $t = t_*$, and thereafter neither recede from nor across our location until $t \rightarrow +\infty$,  thereby ensuring their perpetual convergence towards us. To confirm mathematically this correspondence, we start by addressing the $3^\text{rd}$ scenario of the return journey by treating its condition \eqref{eq77} for zero physical velocity as time approaches infinity, it leads
\begin{equation}
\chi_i = -\lim_{t \rightarrow +\infty} \left( \frac{1}{\dot{a}(t)} \frac{v_0}{\sqrt{a^2(t)(1 - v_0^2) + v_0^2}} \right)
- \int_{0}^{+\infty} \frac{v_0}{a(t') \sqrt{a^2(t')(1 - v_0^2) + v_0^2}} dt'
\label{eq85}
\end{equation}
In our cosmological model, as both $a(t)$ and $\dot{a}(t)$ tend to infinity as $t \rightarrow +\infty$, the limit term in Eq. \eqref{eq85} vanishes which yields the access condition for our latest visitors \eqref{eq84} when $t \rightarrow +\infty$. Consequently, this implies to the following assertion
\begin{align}
3^\text{rd} \ \text{scenario for RJ}: \ \ &\dot{\chi}_{\text{phys}}(+\infty, \chi_i, v_0) = 0\nonumber\\
&\ \ \ \ \ \ \ \ \ \ \ \ \Updownarrow\nonumber\\
\text{Latest visitors}: \ \  &\chi_{\text{phys}}(+\infty, \chi_i, v_0) = 0
\label{eq86}
\end{align}
This equation confirms our proposition that no free particle maintains a constant physical distance, $\dot{\chi}_{\text{phys}}(+\infty) = 0$ relative to our reference frame unless it is perpetually converging towards us, $\chi_{\text{phys}}(+\infty) = 0$, and vice versa. This assertion remains valid regardless of the initial time $t_i$ considered in our analysis. It is worth to mention that the whole analysis might be slightly modified in the case where modified gravity is presented. For a detailed discussion on deviations from the $\Lambda$CDM model, see Ref. \cite{bib19}.
\section{Conclusion}\label{sec8}
In this work, we have presented a complete study of the behavior of radial free-geodesics in the FLRW spacetime based on the most recent  Planck data of $\Lambda$CDM model. In this framework, geodesics are fully determined by two appropriately chosen initial conditions $(\chi_i, v_i)$ and $(\chi_i, v_0)$. Our investigation into geodesic motions across all peculiar velocity regimes has addressed critical questions: How do free test particles behave relative to our reference frame? Which particles have reached, will reach, or will never reach us? These aspects were specifically discussed, and we ultimately concluded that radial free motion in our spacetime follows four potential physical paths:
\begin{enumerate}[leftmargin=*]\addtocounter{enumi}{-1}
\item One-way journey geodesics: Some particles continue moving away from us indefinitely.
\item Double-reversing geodesics: Other particles return to approach us, then start again to move away forever. This phenomenon only occurs after the onset of the universe's accelerated expansion, and these are the only free particles that change their physical path twice.
\item Recrossing geodesics: Some particles return to approach and cross our location.
\item Perpetually approaching geodesics: Finally, there are those particles that approach us indefinitely but never cross our location.
\end{enumerate}
Notably, no free particle remains at a constant physical distance relative to our reference frame unless it is perpetually approaching us through the $3^\text{rd}$ scenario. We have determined the initial conditions for all these scenarios, including the access conditions for particles that reached us, will reach us in the future, or will never reach us, from the Big Bang singularity. This study is crucial for understanding the characteristics of our spacetime and its effect on free particles over time. We aim to extend this investigation to include the gravitational influence of galaxies and the Earth's peculiar velocity relative to the CMB, by considering relativistic velocity transformation in our expanding universe.

\end{document}